\documentstyle[multicol,aps,epsf]{revtex}
\begin{document}

\title{Entropy and Information in Neural Spike Trains}
\author{Steven P. Strong,$^1$ Roland Koberle,$^{1,2}$
Rob R. de Ruyter van Steveninck,$^1$ and William Bialek$^1$}
\address{$^1$NEC Research Institute, 4 Independence Way, Princeton, New
Jersey 08540\\
$^2$Department of Physics, Princeton University, Princeton, New Jersey
08544}

\date{\today}
\maketitle

\begin{abstract}
The nervous system represents time dependent signals in sequences of
discrete action potentials or spikes; all spikes are identical
so that information is carried only in the spike arrival times.
We show how to quantify this information, in bits, free from any
assumptions about which features of the spike train or input signal
are most important, and we
apply this approach
to the analysis of experiments on a motion sensitive neuron in the
fly visual system. This neuron transmits information about the visual stimulus,
at rates of up to 90 bits/s, within a factor of two of the physical
limit set by the entropy of the spike train itself.
\end{abstract}

\begin{multicols}{2}

As you read this text, optical signals reaching your retina are encoded
into sequences of identical pulses, termed action potentials or spikes,
that propagate along the $\sim 10^6$
fibers of the optic nerve from eye to brain.  This
spike encoding appears almost universal, occurring
in animals as diverse as worms and humans, and spanning all the sensory
modalities \cite{adrian}.  The molecular mechanisms for 
the generation and propagation of action potentials are well
 understood \cite{molecular}, as are the mathematical reasons for
the selection of stereotyped pulses by the dynamics of
the nerve cell membrane \cite{mathematical}.  Less well understood
is the function of these spikes as a code \cite{codebook}:
How do the sequences
of spikes represent the sensory world,
and how much information is conveyed in this representation?

The temporal sequence of spikes provides
a large capacity for transmitting information, as emphasized
by MacKay and McCulloch 45 years ago \cite{MM}.
One central question in studies of the nervous system is whether 
the brain takes advantage of this large
capacity, or whether variations in spike timing represent
noise  which must be averaged away \cite{codebook,controversy}.
In response to a long sample of time varying stimuli, the spike train of
a single neuron varies, and we
can quantify this variability by
the entropy per unit time of the spike train, ${\cal S}(\Delta\tau)$
\cite{shannon}, which depends on the time resolution $\Delta\tau$
with which we record the spike arrival times.
If we repeat the same time dependent stimulus, we see a similar,
but not precisely identical, sequence of spikes (Fig. 1).
This variability at fixed input can also be
quantified by an entropy, which we call the conditional or
noise entropy per unit time ${\cal N}(\Delta\tau)$.
The information that the spike train provides about the stimulus
is the difference between the total spike train entropy and this
conditional entropy, $R_{\rm info} = {\cal S} - {\cal N}$ \cite{shannon}.
Because the noise entropy is positive (semi)definite,
the entropy rate sets the capacity for transmitting information,
and we can define an efficiency
$\epsilon (\Delta\tau ) = R_{\rm info} (\Delta\tau )/ {\cal S}
(\Delta\tau)$ with which this capacity is used 
\cite{europhy}.
The question of whether spike timing is important is really the question
of whether this efficiency is high at small $\Delta\tau$
\cite{codebook}.

For some neurons, we understand enough about what the spike
train represents that direct ``decoding'' of the spike train is
possible;
the information extracted by these decoding methods
can be more than half of the total spike train
entropy with $\Delta\tau \sim 1$ ms \cite{europhy}.
The idea that sensory neurons provide a maximally efficient
representation of the outside world has also been suggested as an
optimization principle from which many features of these cells'
responses can be derived \cite{atick}.
But, particularly
in the central nervous system \cite{controversy},
assumptions about what is being encoded should be viewed with
caution. The goal of this paper is, therefore, to give a completely model
independent estimate of entropy and information in neural spike trains
as they encode dynamic signals.

We begin by discretizing the spike train into time bins of size
$\Delta\tau$, and examining segments of the spike train
in windows of length $T$,
so that each possible neural response is a ``word'' with $T/\Delta\tau$ symbols.
Let us call the normalized count of the $i^{\rm th}$ word
${\tilde p}_i$, and then the ``naive estimate'' of the entropy is
\begin{equation}
S_{\rm naive} (T, \Delta\tau ; {\rm size})
= - \sum_i {\tilde p}_i\log_2 {\tilde p}_i ,
\end{equation}
where the notation reminds us that our estimate of
the entropy depends 
on the ${\rm size}$ of the  data set we use in accumulating
the histogram.  The true entropy is
\begin{equation}
S(T,\Delta\tau ) = \lim_{{\rm size}\rightarrow\infty}
S_{\rm naive} (T, \Delta\tau ; {\rm size}) ,
\end{equation}
and we are interested in the entropy rate
\begin{equation}
{\cal S} (\Delta\tau) = \lim_{T\rightarrow\infty} {{S(T,\Delta\tau
)}\over{T}} .
\end{equation}
The difficutly is that, especially at large $T$, very large data sets
are required to ensure convergence of $S_{\rm naive}$ to the true
entropy $S$.

Imagine that we have a
spike train with mean spike rate $\bar r \sim 40$
spikes/s and we sample with a
time resolution $\Delta\tau = 3$ ms.  In a window
of $T= 100$ ms, the maximum entropy 
consistent with this mean rate \cite{codebook,MM} is
$S\sim 17.8\, {\rm bits}$, and we will see that the
entropy of real spike trains is not far from this bound.
But then there are roughly $2^S \sim 2\times 10^5$ words
with significant $p_i$, and if our naive estimation
procedure is going to work, we need to
have at least one sample of each word.
If our samples come from nonoverlapping 100 ms windows,
then we need more than three hours of data,  and one might think that we
need {\it much} more data than this to insure that the probability
of each word is estimated with reasonable accuracy.
Such large quantities of data are generally inaccessible for experiments
on real neurons.

Here we report that it is possible to make progress despite
these pessimistic
estimates.  
First, we examine explicitly the dependence of our entropy estimates on
the size of the data set and find regular behaviors \cite{TP} that
can be extrapolated to the infinite data limit.
Second, we evaluate robust upper \cite {shannon}
and lower \cite{Ma} bounds on the entropy
which serve as a check on our extrapolation procedure.
Third,  we are interested in the extensive component of the entropy in
large time windows, and
we find that a clean approach to extensivity is visible before 
sampling problems set in.
Finally, for the neuron studied---the motion sensitive neuron H1
in the fly's visual system---where we can actually collect
many hours of data.

\vspace{-0.3 cm}
\begin{figure}
\narrowtext
\centerline{ \epsfxsize = 2.5in
\epsffile{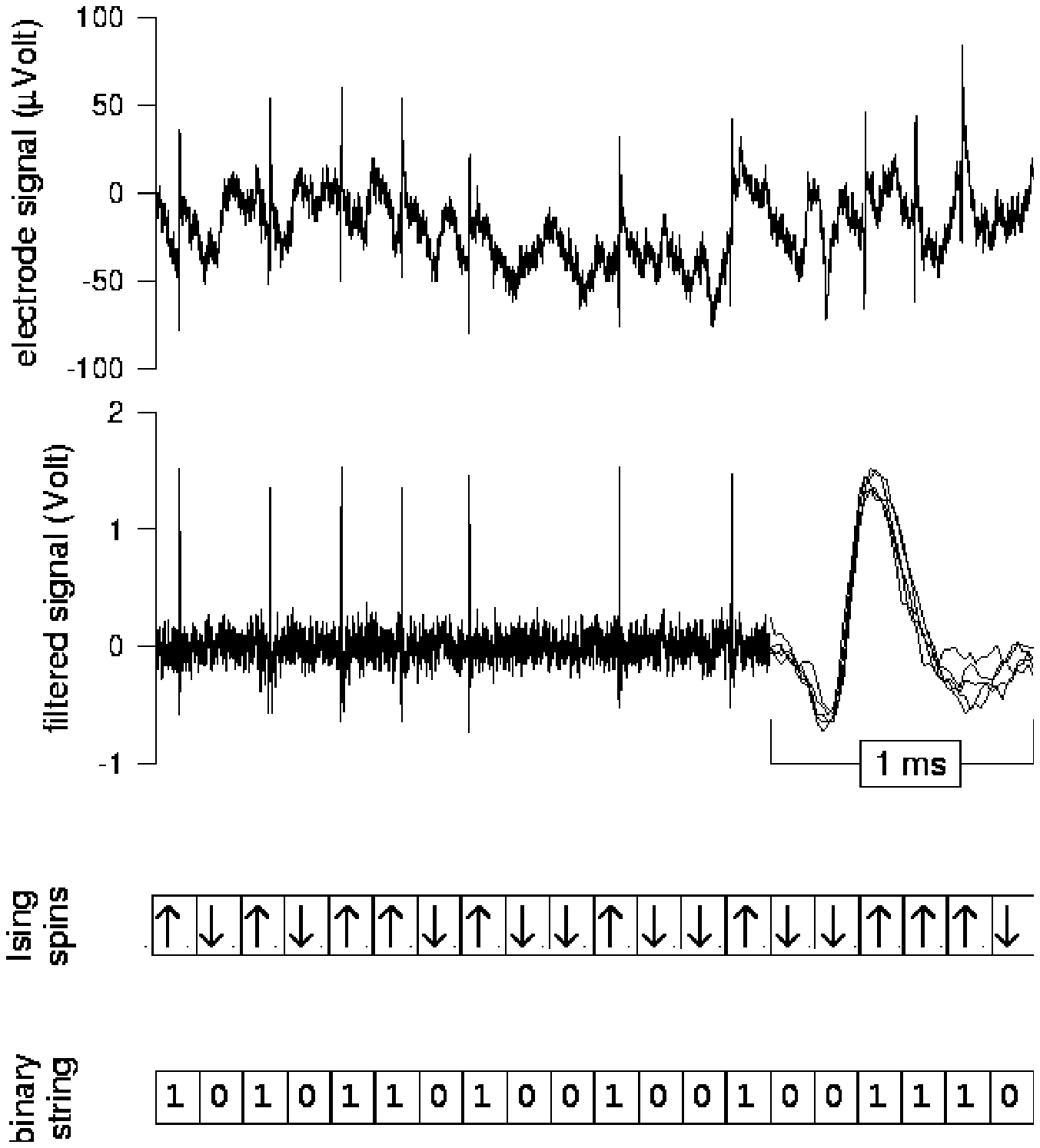}} 
\vspace*{-2.5cm}
\centerline{ \epsfxsize = 2.5in
\epsffile{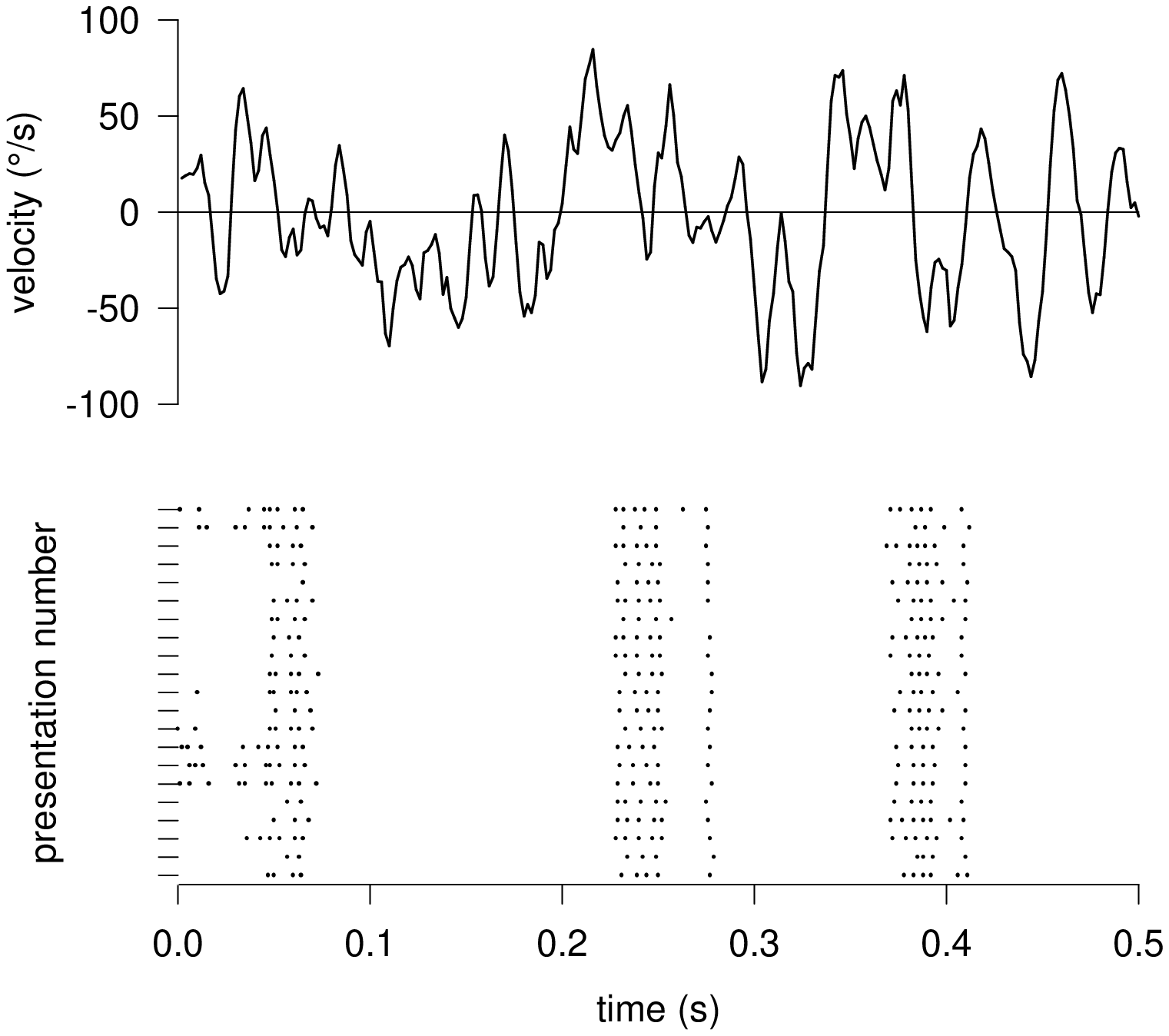}}
\vspace*{-2.5cm}
\caption{
(A) Raw voltage records from a tungsten microelectrode near the cell H1
are filtered and discretized.
(B) Angular velocity of a pattern moving across the fly's visual field
produces a sequence of spikes in H1, indicated by dots. Repeated
presentations produce slightly different spike sequences.
For experimental methods see Ref. [13].}
\end{figure}

H1 responds to motion across the entire visual field, producing more
spikes for an inward horizontal motion and fewer spikes for an outward
motion; vertical motions have no effect on this cell, but are coded by
other neurons \cite{anatomy}.
These cells provide
visual feedback for flight control.  In the experiments analyzed here
\cite{methods},
the fly is immobilized and views computer generated images on a display
oscilloscope.
For simplicity these images consist of a
fixed pattern of vertical stripes with randomly chosen grey levels, and
this pattern takes a random walk in the horizontal direction \cite{stimuli}.

\vspace{-0.3 cm}
\begin{figure}
\narrowtext
\centerline{ \epsfxsize = 2.5in
\epsffile{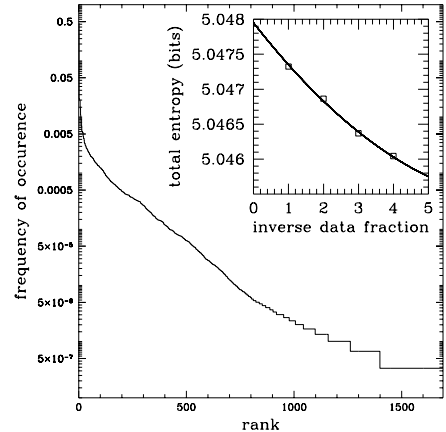}}
\caption{The frequency of occurence for different ``words'' in
the spike train, with $\Delta\tau =3\,{\rm ms}$ and $T=30\,{\rm ms}$.
Words are placed in order so that the histogram is monotonically
decreasing; at this value of $T$ the most likely word
corresponds to no spikes.  Inset shows the dependence
of the entropy, computed from this histogram according to Eq. (1) on
the
fraction of data included in the analysis.
Also
plotted is a least squares fit to
the form $S = S_0  + S_1/{\rm size} + S_2 /{\rm size}^2$.
The intercept $S_0$
is our extrapolation to the true value
of the entropy with infinite data [10].}
\end{figure}

We begin our analysis with time bins of size $\Delta\tau =$ 3 ms.
For a window  of $T= 30$ ms---corresponding to the behavioral
response time of the fly \cite{30ms}---we can estimate
the entropy rather accurately by the
naive procedure described above.  Figure 2 shows the histogram
$\{{\tilde p}_i\}$, and the naive entropy estimates.
We see
that there are very small finite data set corrections ($< 10^{-3}$),
well fit by
\cite{TP}
\begin{eqnarray}
S_{\rm naive} (T, \Delta\tau ; {\rm size})
&=& S (T , \Delta\tau)  + {S_1(T , \Delta\tau) \over{\rm size}}
\nonumber\\
&&\,\,\,\,\, + {{S_2 (T, \Delta\tau)}\over {{\rm size}^2}} .
\end{eqnarray}
Under these conditions we feel confident that the extrapolated
$S (T, \Delta\tau)$ is the correct entropy.
For sufficiently large $T$, sampling problems occur:
finite size corrections become much larger,
the contribution of the second correction is significant
and the extrapolation to infinite size is
unreliable.

Ma \cite{Ma} discussed the problem of entropy estimation in the
undersampled limit.  For probability distributions that are uniform on a
set of $N$ bins (as in the microcanonical ensemble),
the entropy is $\log_2 N$ and the problem is to
estimate $N$.  Ma noted that this could be done by counting the
number of times that two randomly chosen observations yield
the same configuration, since the probability of such a coincidence is
$1/N$.
More generally, 
the probability of a coincidence is $P_{\rm c} = \sum p_i^2$,
and hence
\begin{eqnarray}
S &=& -\sum p_i \log_2 p_i  = - \langle \log_2 p_i \rangle \nonumber\\
&\geq& -\log_2 \left( \langle p_i \rangle \right) = -\log_2 P_c ,
\label{bound}
\end{eqnarray}
so we can compute a lower bound the the entropy by counting
coincidences. Furthermore, as emphasized by Ma,
$\log_2 P_c$ is less sensitive to sampling errors
than is $S_{\rm naive}$.  The Ma bound is the minimum entropy consistent with a
given $P_c$, and it is one of the Renyi entropies \cite{Renyi}.
It is also at the heart of algorithms
for the analysis of attractors in dynamical systems \cite{ruelle}.

The Ma bound is tightest for distributions
that are close to uniform.  The distributions of neural
responses cannot be uniform because the spikes are sparse.  But
the distribution of words with fixed spike count, $N_{\rm sp}$,
is  more nearly uniform, so we apply the Ma bounding procedure
in each $N_{\rm sp}$ sector.  Thus $S\geq S_{\rm Ma}$,
with
\begin{eqnarray}
\label{eq:mabound}
S_{\rm Ma} &=& - \sum_{N_{\rm sp}} P(N_{\rm sp})
\nonumber\\
&&\,\,\,\,\times
\log_2 \left[ P(N_{\rm sp})
\frac{2n_c(N_{\rm sp})}{N_{\rm obs}(N_{\rm sp})[N_{\rm
obs}(N_{\rm sp})-1]}
\right] ,
\end{eqnarray}
where $n_c(N_{\rm sp})$ is the number of coincidences
observed among the words with $N_{\rm sp}$ spikes,
$N_{\rm obs}(N_{\rm sp})$ is the total number of occurrences of words
with $N_{\rm sp}$ spikes,
and $ P(N_{\rm sp})$ is the fraction of words with
$N_{\rm sp}$ spikes.

In Fig. 3 we plot the
entropy  as a function of the window size $T$, with results both from
the naive procedure and from the Ma bound. 
For sufficiently large windows the naive procedure gives an answer
smaller than the Ma bound, and hence the naive answer must be unreliable
because it is more sensitive to sampling problems.  
Before this sampling disaster 
the lower bound and the naive estimate are
never more than  10--15\% apart.
The point at which the naive estimate
crashes into the Ma bound is also where the second
correction in Eq. (4) becomes significant and we lose control over the
extrapolation to the infinite data limit.
This occurs at $T\sim 100\ {\rm  ms}$.
We can trust the Ma bound beyond
this point, but it becomes steadily less powerful.

\vspace{-0.3 cm}
\begin{figure}
\narrowtext
\centerline{ \epsfxsize = 2.5in
\epsffile{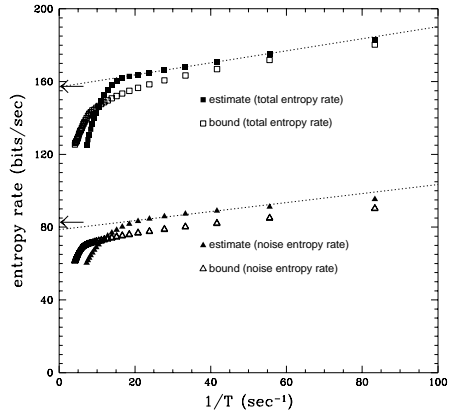}}
\caption{
The total and noise entropies
per unit time (in bits per second) are plotted  versus the
reciprocal of the window size (in s$^{-1}$), with the time resolution
held fixed at $\Delta\tau =3$ ms.
Results are given both for the direct estimate and for the bounding
procedure described in the text, and for each data point
we apply the extrapolation procedures of Fig. 2 (inset).
Dashed lines indicate extrapolations to infinite word length, as
discussed
in the text, and
arrows indicate upper bounds obtained by differentiating $S(T)$
[7].}
\end{figure}

If the correlations in the spike train have finite range, then 
the leading subextensive contribution to the entropy will be a
constant $C(\Delta\tau)$.  This means that
\begin{equation}
{{S(T,\Delta\tau)}\over{T}}
= {\cal S} (\Delta\tau ) + {{C(\Delta\tau)}\over T} + \cdots .
\end{equation}
This asymptotic behavior is seen
clearly in Fig. 3, and emerges before the sampling disaster. 
Given the clean linear behavior
in a well sampled region of the plot, we trust the extrapolation and
arrive at an estimate of the entropy per unit time,
${\cal S} (\Delta\tau = 3\,{\rm ms}) =157 \pm 3$ bits/s. 

The entropy approaches its extensive limit from above
\cite{shannon}, so that
\begin{equation}
{1\over {\Delta\tau}}\left[ S(T + \Delta\tau , \Delta\tau )
- S(T, \Delta\tau)\right]  \geq {\cal S}
(\Delta\tau)
\end{equation}
for all window sizes $T$.
This bound becomes progressively tighter at larger $T$, until
sampling problems set in. In fact there is a broad plateau
($\pm 2.7\%$) in the range $18 < T < 60$ ms, leading to 
$S \leq 157\pm 4$ bits/s, in excellent agreement with the extrapolation
in Fig. 3.  

The noise entropy per unit time ${\cal N}(\Delta \tau)$
measures the variability of the spike train when the input signals are held 
fixed.
Hence we need to look at the ensemble of responses
to repeated presentations of the same time varying input signal.
Marking a particular time $t$ relative to the stimulus,
we accumulate the frequencies of occurrence of each word $i$ that
begins at $t$, and call this histogram $\tilde p_i (t)$.
This generates a naive estimate of the local noise entropy in the window
from $t$ to $t+T$,
\begin{equation}
N_{\rm naive}^{\rm local} (t,T,\Delta\tau; {\rm size}) =
-\sum_i \tilde p_i (t) \log_2 \tilde p_i (t) .
\end{equation}
Estimating the average rate of information transmission by the spike
train requires knowing the noise entropy averaged over $t$,
$N_{\rm naive} (T,\Delta\tau; {\rm size}) = \langle N_{\rm naive}^{\rm
local} (t;T,\Delta\tau, {\rm size})\rangle_t$.  Then we analyze as
before the extrapolation to large data set ${\rm size}$ and large $T$.
Fig. 3 also shows the noise entropy results
as a function of the window size.
The difference between the two
entropies is the information which the cell transmits,
$R_{\rm info} (\Delta\tau = 3\,{\rm ms}) = 78 \pm 5$
bits/s, or $1.8 \pm 0.1$ bits/spike.

This anlysis has been carried out over a range of time resolutions,
$800 > \Delta\tau > 2$ ms, and the results are summarized in Fig. 4.
Over this range, the entropy per unit time of the spike train varies over
a factor of roughly 40, illustrating the increasing capacity of the
system to convey information by making use of spike timing.  Note
that $\Delta\tau = 800$ ms corresponds to counting spikes in bins that
contain typically thirty spikes, while $\Delta\tau =2$ ms
corresponds to timing each spike to within 5\% of the typical interspike
interval.  The information that the spike train conveys
about the dynamics of motion across the visual field increases 
in approximate  proportion to the entropy, corresponding to 
$\sim 50\%$ efficiency, at least for this ensemble of stimuli.

\vspace{-0.3 cm}
\begin{figure}
\narrowtext
\centerline{ \epsfxsize = 2.5in
\epsffile{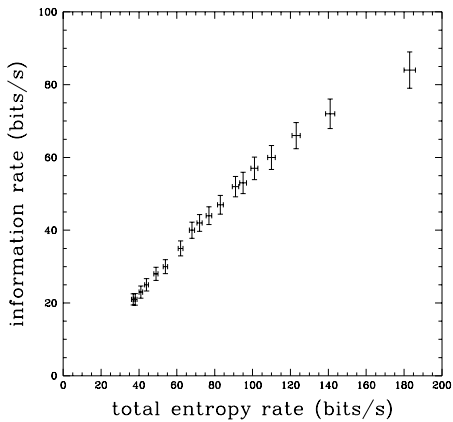}}
\caption{Information and entropy rates computed at various time
resolutions, from $\Delta \tau = 800\,{\rm ms}$ (lower left)
to $\Delta\tau = 2\,{\rm ms}$ (upper right).  Note the approximately
constant, high efficiency across the wide range of entropies.}
\end{figure}

Although we understand a good deal about the signals
represented in H1 \cite{anatomy,science}, our present analysis 
does not hinge on this knowledge.
Similarly, although it is possible to collect very large data sets
from H1, Fig.'s 2 and 3 suggest that more limited data sets
would compromise our conclusions only slightly.
It is feasible, then, to apply these same analysis techniques to
cells in the mammalian brain \cite{mt}.
Like cells in the monkey or cat primary visual cortex,
H1 is four layers
`back' from the array of photodetectors and receives its inputs from
thousands of synapses.
For this central neuron, half the available information capacity is
used, down to millisecond precision.
Thus, the analysis developed here allows us for the first time to
demonstrate directly the significance of spike timing in the nervous system
without any hypotheses about the structure of the neural code.

We thank N. Brenner,  K. Miller, and P. Mitra for helpful discussions, and
G. Lewen for his help with the experiments.  R. K., on leave
from the Institute of Physics, Universidade de S\~ao Paulo,
S\~ao Carlos, Brasil, was supported in part by CNPq and FAPESP.

\end{multicols}

\begin{thebibliography}{99}
\bibitem{stimuli}
The details of entropy and information in spike trains obviously depend
on stimulus conditions.  In these experiments the fly is immobilized
13.24 cm from the screen of a display oscilloscope and views
images with an angular area of 875 deg$^2$.  The photoreceptor
cells in the fly form a regular lattice, so we can observe aliasing
in the response of the movement sensitive neurons and hence measure the
lattice constants; we then calculate that this
image area stimulates 3762 photoreceptors.  The mean radiance of the
screen is 180 mW/(sr$\cdot$m$^2$), and from the receptor geometry
and other experiments this corresponds to a photon counting rate of
$5\times10^4\,{\rm s}^{-1}$ in each photoreceptor.  This is, roughly, the light 
level
for a fly flying at dusk.  The pattern on the screen consists of bars
randomly set dark or light, with the bar width equal to the receptor
lattice spacing.  Finally, the pattern takes a random walk in 2 ms time
steps, corresponding to the refresh time of the monitor; the 
diffusion constant is 2.8 deg$^2$/s.
\end{thebibliography}
\end{document}